\documentclass{nature}
\usepackage{bm}
\usepackage{graphicx}

\title{Heat Superconductivity}

\author{Junren Shi$^{1}$, \& Zhengqian Cheng$^1$}

\begin{document}

\maketitle

\begin{affiliations}
\item International Center for Quantum Materials, Peking University, Beijing 100871, China
\end{affiliations}

\begin{abstract}
Electrons/atoms can flow without dissipation at low temperature in superconductors/superfluids. The phenomenon known as superconductivity/superfluidity is one of the most important discoveries of modern physics, and is not only fundamentally important, but also essential for many real applications~\cite{isbn9780070648784}. An interesting question is: can we have a superconductor for heat current, in which heat/energy can flow without dissipation? Here we show that heat superconductivity is indeed possible.  We will show how the possibility of the heat superconductivity emerges in theory, and how the heat superconductor can be constructed using recently proposed time crystals~\cite{wilczek2012,li2012}. The underlying simple physics is also illustrated (see discussions after Eq.~(\ref{supercurrent})). If the possibility could be realised, it would not be difficult to speculate various potential applications, from heat tele-transportation to cooling of information devices.
\end{abstract}

The usual superconductivity emerges as a result of spontaneous breaking of electromagnetic gauge symmetry~\cite{isbn9780201328301}. The gauge symmetry dictates that all electron systems are invariant under changes of phase pre-factors of wave amplitudes of constituent electrons. The conservation of electron numbers is a result of the symmetry, according to Noether's theorem. A superconductor breaks the gauge symmetry: its ground state is not a phase symmetric and fixed-particle-number state, as expected from the gauge symmetry.  Instead, a macroscopic superconductor has definite phase for its constituent particles (Cooper pairs).  When coupling two superconductors with different phases, one finds a super-current between them, known as Josephson effect~\cite{isbn9780070648784}. More generally, a slow spatial variation of the phase in continuous space will drive a super-current.  

The above discussion provides useful hints on how the heat superconductivity could emerge. Like the electron numbers, energy is a conserved quantity. Actually, it is the only quantity that is conserved for all imaginable physical systems. The symmetry associating with the conservation of energy is nothing but the time translational symmetry.  An analogy to the usual superconductivity immediately suggests that the heat superconductivity could emerge as a result of spontaneous breaking of the time-translational symmetry, which is equivalent to spontaneous emergence of local clocks~\cite{wilczek2012}.  One would expect a spatial variation of local clocks could drive a heat super-current, i.e., the Josephson effect for heat. 

The time-crystal proposed by Wilczek et al. provides an instance of spontaneous time translational symmetry breaking~\cite{wilczek2012,shapere2012a}. A couple of model systems to realise the time crystal had been proposed, including soliton phase of a collection of charge particles with attractive interaction confined in a ring~\cite{wilczek2012}, and Wigner crystal phase of a group of ultracold ions confined in a ring-shaped trapping potential~\cite{li2012}.  In both cases, the system undergoes spontaneous symmetry breaking of spatial translational symmetry, and forms a phase of inhomogeneous density distribution. As a result, there emerges a new parameter $\theta_0$, which specifies center of the soliton or the origin of the Wigner crystal.  In the macroscopic limit of the particle number $N\rightarrow \infty$, there is emergent orthogonality between different $\theta_0$-states, and for all local observations, the system could be considered to be in a state with definite $\theta_0$~\cite{wilczek2012}.

A magnetic field transforms the system to a time crystal, which can be visualised as a persistent rotator with a finite angular velocity in its ground state~\cite{wilczek2012,li2012}:
\begin{equation}
\omega_0 = (l +\alpha) \frac{\hbar}{I} \, ,
\end{equation}
where $\alpha = q\phi/h$ with $\phi$ being the total magnetic flux threading through the ring, $q$ is charge of ion/particle, $I$ is momentum of inertia of an ion/particle relative to the center of the ring, and $l$ is an integer (or a half-integer for an odd number of fermions) that minimises $|l +\alpha|$. The corresponding ground state wave function can be related to the zero magnetic field $\theta_0$-state $\Psi(\{\theta_{i}\}, \theta_0)$ by~\cite{wilczek2012}:
\begin{equation}
\Psi^{\alpha}(\{\theta_{i}\},\theta_0)=e^{-\mathrm{i}l\sum_{i}\theta_{i}}\Psi(\{\theta_{i}\}, \theta_0 -\omega_0 t)\, , \label{wftc}
\end{equation}
where $\{\theta_{i}\}$ denotes the set of coordinates of particles on the ring.  When $\alpha$ is not an integer or a half-integer~\cite{wilczek2012}, the ground-state many-body wave-function becomes time-dependent, and the system becomes a persistently rotating object. Here, $\theta_0$ acquires a new meaning: the phase angle of the persistent rotation. Effectively, it defines the local clock.  

\begin{figure}
\begin{center}
\includegraphics[width=0.7\textwidth]{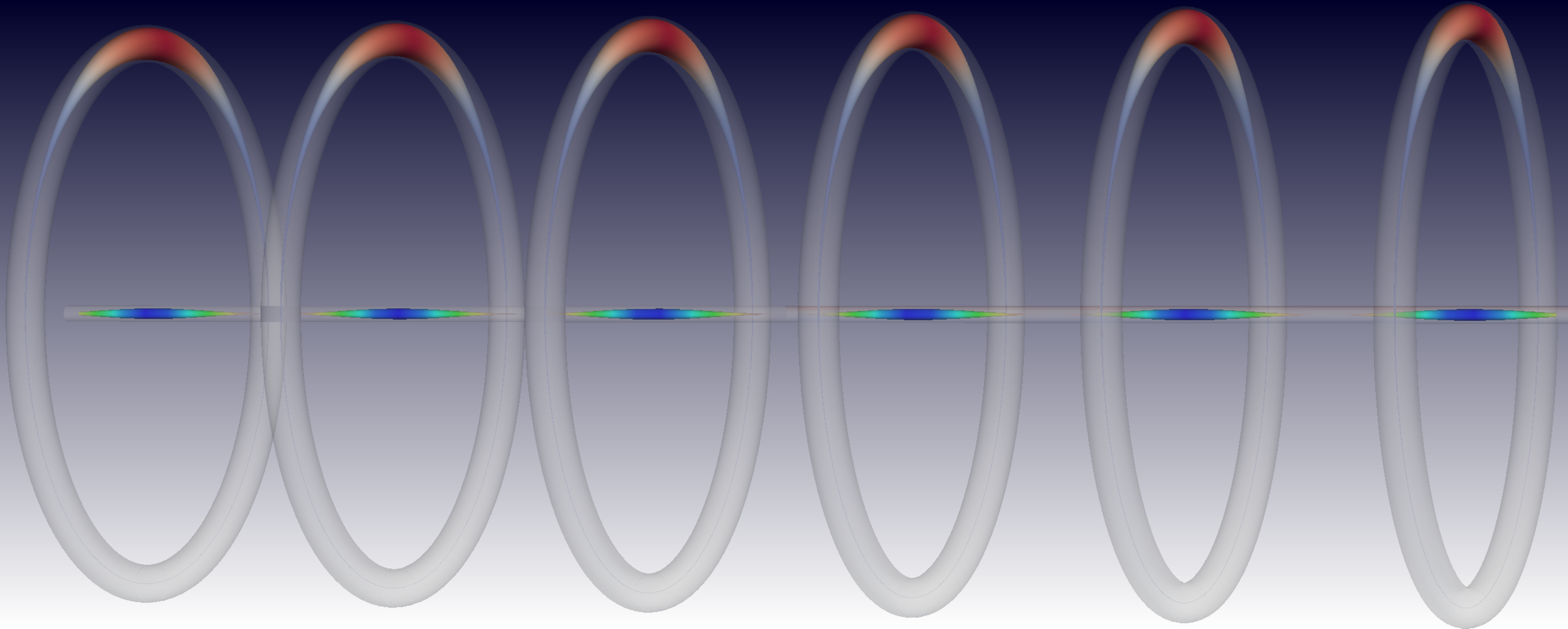}

{\bf Figure 1 (a)}


\includegraphics[width=0.7\textwidth]{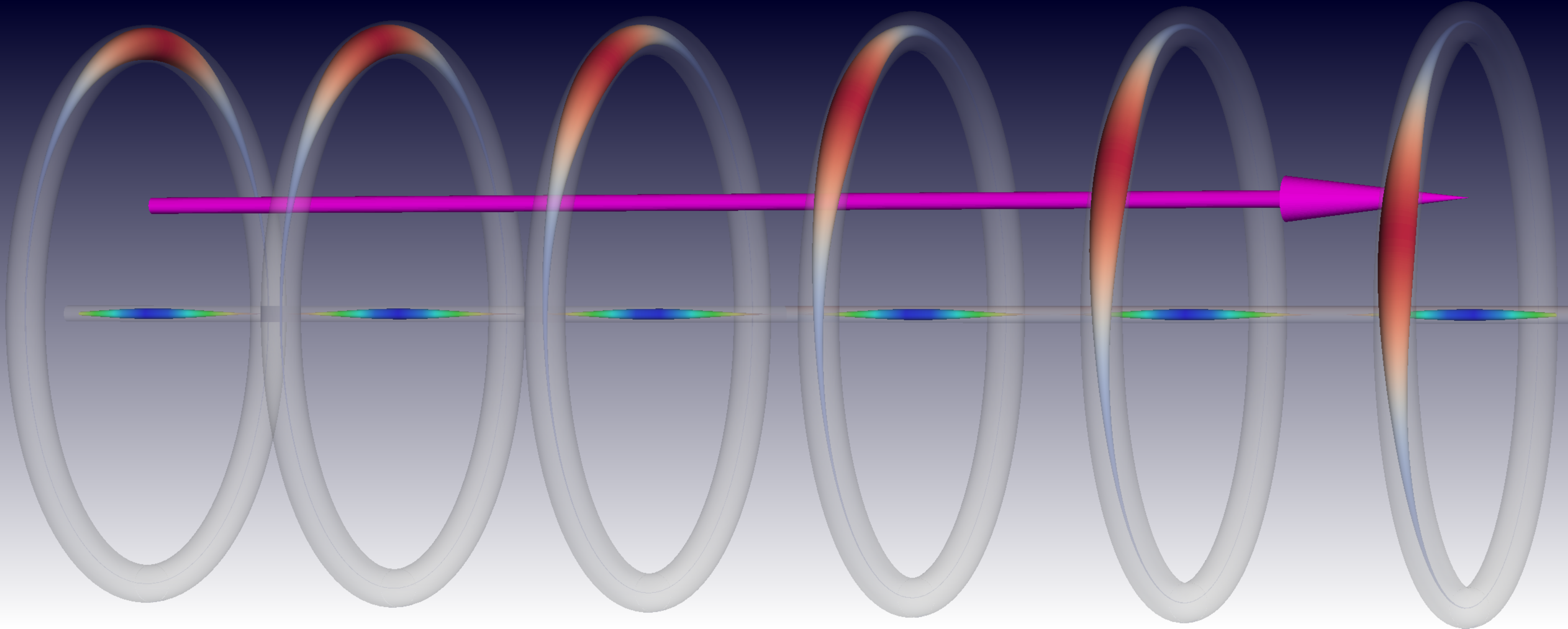}

{\bf Figure 1 (b)}
\end{center}
\caption{\label{WignerRings} (a) The structure of a heat superconducting rod. A series of soliton rings/Wigner crystal rings are arranged concentrically along the direction perpendicular to the rings. The rings are stabilised to a configuration of relative phase angles by the inter-ring couplings.  All rings rotate synchronously and persistently at angular velocity $\omega_0$. A conducting filament is threaded through the central axis of the rod and in contact with external electrodes, providing free charges to compensate the charges trapped on the rings. Here, the particle density profiles of soliton rings are shown and colour-coded.  The spacings between rings are exaggerated.  (b) A twist of the rod induces a heat super-current, indicated by the magenta arrow.}
\end{figure}

A heat superconductor can be constructed from a group of time crystals by stacking them concentrically to form a three-dimensional rod, as shown in Fig.~\ref{WignerRings}. At sufficiently low temperature, the inter-ring coupling due to the density inhomogeneity will stabilise a configuration of relative phase angles of the rings.  It also gives rise to a finite torsional stiffness. The elastic energy induced by a twist of the rod can be written as, 
\begin{equation}
E - E_0 =\frac{\kappa}{2}\int\mathrm{d}x\left[\partial_{x}\theta_0(x)\right]^{2} \, .\label{dE}
\end{equation}
in the continuum limit, where $E_0$ is the ground state energy, $\kappa$ is the torsional stiffness, and we assume that the rod is extended along the $x$-direction.

To determine heat current along the rod, we introduce a gravitational vector potential $A_E(x)$, which couples to the heat current in a way just like that of the usual vector potential to the electric current. We apply the local gravitational gauge symmetry (see Methods): the system should be invariant under joint transformations $(x,t)\rightarrow (x,t+\chi_E(x))$ and $A_E(x) \rightarrow A_E(x) + \partial_x\chi_E(x)$ to the first order of a slowly varying function $\chi_E(x)$.  For the ground state wave function Eq.~(\ref{wftc}) of time crystal, the local time translation is equivalent to a translation of the phase angle $\theta_0(x) \rightarrow \theta_0(x) - \omega_0 \chi_E(x)$.  The symmetry dictates the minimal coupling form to the gravitational vector potential: 
\begin{equation}
E - E_0 =\frac{\kappa}{2}\int\mathrm{d}x\left[\partial_{x}\theta_{0}(x)+\omega_{0}A_{E}(x)\right]^{2} \, . \label{degauge}
\end{equation}

The energy current along the rod can then be determined:
\begin{equation}
j_{E}^{(s)}(x)=-\left.\frac{\delta E}{\delta A_{E}(x)}\right|_{A_{E}(x)\rightarrow0}=-\kappa\omega_{0}\partial_{x}\theta_{0}(x) \, .\label{supercurrent}
\end{equation}
We see that a heat (energy) super-current emerges in a twisted rod in the presence of the persistent rotation $\omega_0$.  The result is consistent to our general theoretical consideration, as a spatial variation of $\theta_0(x)$ implies a variation of the local clocks.

\begin{figure}
\begin{center}
\includegraphics[width=0.7\textwidth]{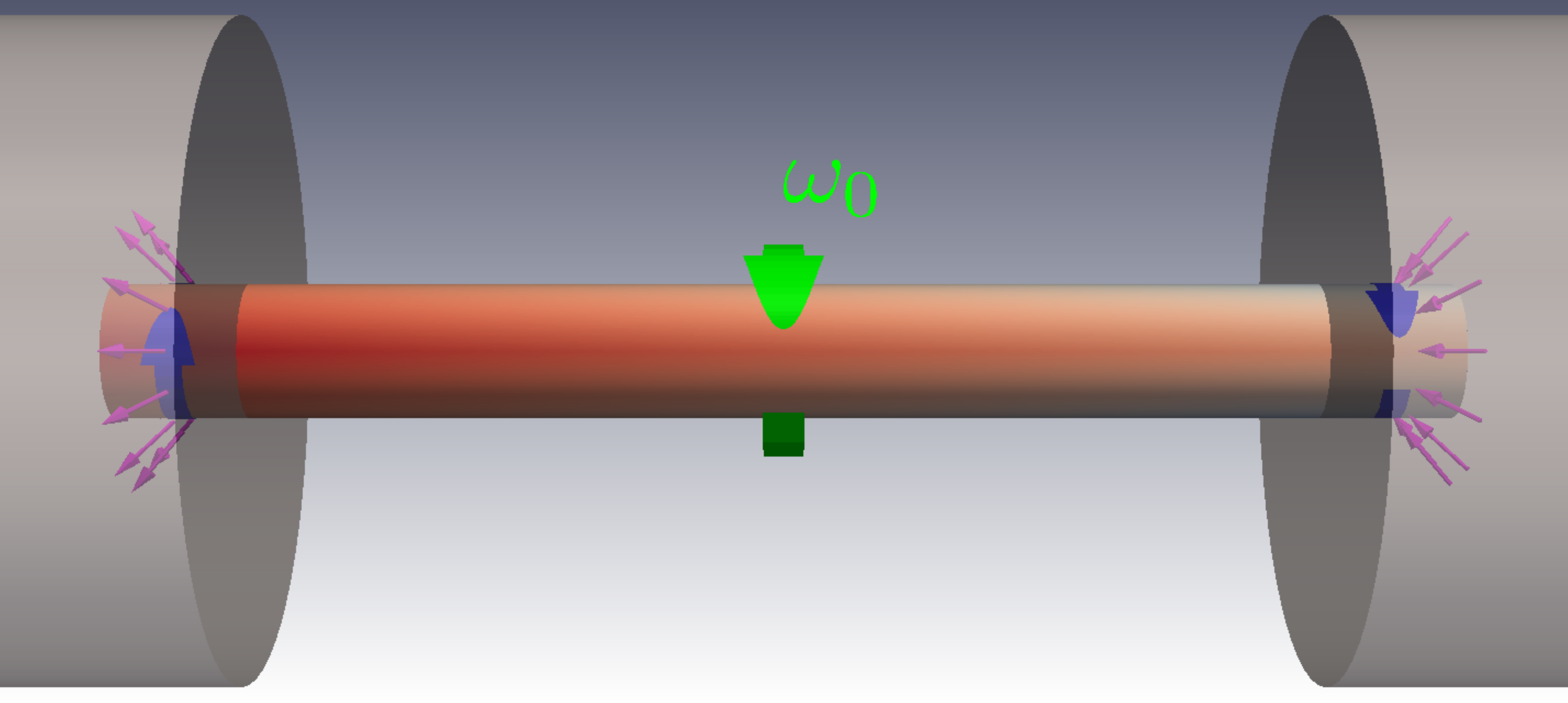}

{\bf Figure 2}
\end{center}
\caption{\label{RotatingCylinder} A heat superconductor in contact with environment. A structure of Fig.~\ref{WignerRings} is assembled and represented by a rod in the middle. The rod has a persistent angular velocity $\omega_0$. The two ends of the rod submerge into two reservoirs of viscosity fluid (represented by two cylinders at the two ends).  The directions of friction torques exerted by the fluid reservoirs to the rod are shown by the blue arrows.  In the presence of both the torques and the persistent rotation, there will be energy exchange between the rod and the two fluid reservoirs. The corresponding directions of the energy fluxes are shown by magenta arrows.  At the stationary state, the friction torques at the two ends must have the same magnitude and the opposite directions. }
\end{figure}

Equation (\ref{supercurrent}) is the central result of this paper. The underlying simple physics is illustrated in Fig.~\ref{RotatingCylinder}.  We consider a heat superconducting rod in contact with environment through two viscosity fluid reservoirs. Had the centre rod been the usual solid, the viscosity fluid would exert friction torque that will eventually put it to rest.  However, a rod made of the time crystals rotates persistently at the angular velocity $\omega_0$, and cannot be slowed down because it is already in the ground state. 

We consider the situation when the environment exerts torques $\mathcal{T}_{L(R)}$ at the two ends of the rod.  In the presence of the persistent rotation, the torques will do works $W_{L(R)}$ to the rod~\cite{isbn9780201657029}:
\begin{equation}
\frac{\mathrm{d} W_{L(R)}}{\mathrm{d} t} =  \omega_0 \mathcal{T}_{L(R)}\, . \label{WLR}
\end{equation}
The required energy must be absorbed from or released to the environment.  At the stationary state that the phase angle configuration of the rod is fixed, there is no net energy exchange between the rod and the environment, and we must have:
\begin{equation}
\left\langle \mathcal{T}_R \right\rangle = - \left\langle \mathcal{T}_L \right\rangle = \mathcal{T} \, .\label{TLR} 
\end{equation}
As a result, the energy will be absorbed from the environment at the one end, and released to the environment at the other end.  Effectively, a heat current $\omega_0 \mathcal{T}$ passes through the rod from one end to the other. At the same time, an average torque $\mathcal{T}$ will induce a twist:
\begin{equation}
\partial_x \theta_0(x) = - \mathcal{T}/\kappa \, . \label{deformation}
\end{equation}
Combining Eq.~(\ref{WLR}) and (\ref{deformation}), we obtain the heat super-current formula Eq.~(\ref{supercurrent}).

The emergence of the torques at the two ends of the rod is resulted from the microscopic process of energy exchanges with the environment.  At the finite temperature, the two ends of the rod continuously receive random torsional kicks from the environment.  Each random kick pushes the system out of its ground state temporarily, accompanying with energy exchanges. The system reaches equilibrium when energy absorptions and emissions are balanced.  For a time crystal rod, the process gives rise to temperature dependence of equilibrium expectation value of the persistent angular velocity $\omega_0(T)$~\cite{li2012}.  When the system deviates from the equilibrium, there will be net energy influx or outflow, corresponding to an average torque:
\begin{equation}
\left\langle\mathcal{T} \right\rangle = \frac{J_E}{\omega_0(T)}\, , \label{torques}
\end{equation}
where $J_E$ is the net energy flux.



We now consider an infinitesimal temperature difference $\delta T$ such that the left (right) reservoirs in Fig.~\ref{RotatingCylinder} has temperature $T_{L(R)} = T \pm \delta T/2$. For a symmetric structure, the rod must be at the temperature $T$, maintaining a persistent angular velocity $\omega_0(T)$.  It is important to note that the whole rod must be at the same temperature so that the different parts of the rod could rotate synchronously.  The temperature differences between the rod and the environment at the two ends will drive net energy fluxes,
\begin{equation}
J_E^{L(R)} = \pm \kappa_T^{SN} \frac{\delta T}{2}\, ,
\end{equation}
where we introduce an interface heat conductance $\kappa_T^{SN}$ for heat conduction between the normal and superconducting heat mediums.  The corresponding torques determined by Eq.~(\ref{torques}) twist the rod, inducing heat super-current in the rod. The calculation of $\kappa_T^{SN}$ depends on details of microscopic model, in analogy to the calculation of the tunnelling conductance of  a normal metal-superconductor junction in the usual superconductivity~\cite{isbn9780070648784}. To transport heat, one does need a temperature difference to overcome the contact heat resistance $2/\kappa_T^{SN}$ of the two interfaces.  On the other hand, the heat flow generated by a given temperature difference does not depend on the length of the central heat superconducting rod, meaning that one can transmit heat for arbitrary distance without incurring transmission loss in the rod. 

There is also AC Josephson effect counterpart in heat superconduction~\cite{isbn9780070648784}. This can be achieved by cutting the centre rod of Fig.~\ref{RotatingCylinder} into two weakly coupled disjointed pieces. The two pieces will have different persistent angular velocities $\omega_0(T_L)$ and $\omega_0(T_R)$ when the temperatures of the two reservoirs are different.  The resulting heat super-current will oscillate at frequency $|\omega_0(T_L) - \omega_0(T_R)|$ or one of its harmonics. 

In summary, we show that the heat superconductivity naturally emerges in a system spontaneously breaking the time translational symmetry.  Although our discussion is based upon the existing models of time crystal and has a mechanical intepretation, the essential physics is completely general, i.e., a spatial variation of local clocks will drive a heat super-current.
The possibility to get a real heat superconductor in laboratory is certainly rested on the realisation of the time crystal, which is still challenging both theoretically and experimentally.  On the other hand, one should not ignore the possibility of finding the heat superconductivity in natural compounds, as the usual superconductivity was first discovered in mercury. Nevertheless, it is yet to see any fundamental principle prohibiting the spontaneous breaking of time translational symmetry from happening, and a connection between it and a new matter of vast application potential (i.e., heat superconductor) would be a great incitation for future pursuits. 

\begin{methods}

Luttinger introduces a gravitational scalar potential $\psi(\bm r)$ which perturbs the hamiltonian~\cite{luttinger1964}:  
\begin{equation}
\hat{H} \rightarrow \int\mathrm{d}\bm r \left[1+\psi(\bm r)\right] \hat{h}(\bm r)\, ,\label{scaleH}
\end{equation}
where $\hat{h}(\bm r)$ is the local hamiltonian density, and $\hat{H} = \int \mathrm{d} \bm{r} \hat{h}(\bm r)$. One can further introduce gravitational vector potential $\bm A_E(\bm r)$ which is coupled to the energy current operator $\hat{\bm j}_E(\bm r)$: 
\begin{equation}
   \hat{H} \rightarrow \int\mathrm{d}\bm r \left[\hat{h}(\bm r) - \hat{\bm j}_E(\bm r) \cdot \bm A_E(\bm r)  \right] \, . \label{AEcoupling}
\end{equation}
The set $(\psi,\, \bm A_E)$ provides heat transport counterparts of electromagnetic scalar and vector potentials~\cite{braginsky1977,ryu2012}. 

The energy current operator $\bm j_E(\bm r)$ is defined by the energy continuity-conservation equation~\cite{qin2011}:
\begin{equation}
\frac{\partial \hat{h}(\bm r)}{\partial t} \equiv \frac{1}{\mathrm{i}\hbar} \left[\hat{h}(\bm r),\, \hat{H}\right] = -\bm\nabla \cdot \hat{\bm j}_E(\bm r) \, .
\end{equation}
The equation only defines the energy current up to a curl.  To eliminate the uncertainty, we further impose the scaling law~\cite{qin2011,cooper1997,qin2012}:
\begin{equation}
\hat{\bm j}_E(\bm r) \rightarrow \left[1+\psi(\bm r)\right]^2 \hat{\bm j}_E(\bm r) \, \label{scaling} 
\end{equation} 
when the hamiltonian is scaled by the gravitational scalar potential as Eq.~(\ref{scaleH}).

We can introduce the local gravitational gauge transformation:
\begin{equation}
\hat{U}_{\mathrm{G}} = \exp\left[-(\mathrm{i}/\hbar)\int \mathrm{d}\bm r \chi_E(\bm r)\hat{h}(\bm r)\right]\, ,
\end{equation}
where $\chi_E(\bm r)$ is a slowly varying function.  It is easy to see that the transformation is just a local time translation $(\bm r, t) \rightarrow (\bm r,\, t+\chi_E(\bm r))$ if one ignores the gradient corrections of $\chi_E(\bm x)$.

Under the local gravitational gauge transformation, the local hamiltonian density is transformed, to the first order of $\chi_E(\bm r)$:
\begin{equation}
U_{\mathrm{G}}\hat{h}(\bm r)U_{\mathrm{G}}^{\dagger}\approx \hat{h}(\bm r) - \frac{1}{\mathrm{i}\hbar} \left[\hat{h}(\bm r),\,  \int \mathrm{d}\bm r \chi_E(\bm r) \hat{h} (\bm r) \right] =
\hat{h}(\bm r)+\bm\nabla\chi_E \cdot \hat{\bm j}_{E}(\bm r)+\bm\nabla\cdot\left[ \chi_E\hat{\bm j}_{E}(\bm r) \right]\, . \label{eq:H-trans}
\end{equation}
The commutator can be derived from the scaling law Eq.~(\ref{scaling})~\cite{qin2011}.


Using Eq.~(\ref{eq:H-trans}), it is easy to verify the local gravitational gauge symmetry: to the first order of $\chi_E(\bm r)$, the total hamiltonian $\hat{H}$ is invariant under the joint transformations $\hat{U}_{\mathrm{G}}$ and $\bm A_E(\bm r) \rightarrow \bm A_E(\bm r) + \bm\nabla \chi_E(\bm r)$.

\end{methods}



\begin{thebibliography}{10}
\expandafter\ifx\csname url\endcsname\relax
  \def\url#1{\texttt{#1}}\fi
\expandafter\ifx\csname urlprefix\endcsname\relax\def\urlprefix{URL }\fi
\providecommand{\bibinfo}[2]{#2}
\providecommand{\eprint}[2][]{\url{#2}}

\bibitem{isbn9780070648784}
\bibinfo{author}{Tinkham, M.}
\newblock \emph{\bibinfo{title}{Introduction to superconductivity}} (\bibinfo{publisher}{McGraw Hill},
  \bibinfo{year}{1996}).

\bibitem{wilczek2012}
\bibinfo{author}{Wilczek, F.}
\newblock \bibinfo{title}{Quantum time crystals}.
\newblock \emph{\bibinfo{journal}{Phys. Rev. Lett.}}
  \textbf{\bibinfo{volume}{109}}, \bibinfo{pages}{160401}
  (\bibinfo{year}{2012}).

\bibitem{li2012}
\bibinfo{author}{Li, T.} \emph{et~al.}
\newblock \bibinfo{title}{{Space-Time} crystals of trapped ions}.
\newblock \emph{\bibinfo{journal}{Phys. Rev. Lett.}}
  \textbf{\bibinfo{volume}{109}}, \bibinfo{pages}{163001}
  (\bibinfo{year}{2012}).

\bibitem{isbn9780201328301}
\bibinfo{author}{Anderson, P.~W.}
\newblock \emph{\bibinfo{title}{Basic notions of condensed matter physics}}
  (\bibinfo{publisher}{Addison-Wesley}, \bibinfo{year}{2010}).
  

\bibitem{shapere2012a}
\bibinfo{author}{Shapere, A.} \& \bibinfo{author}{Wilczek, F.}
\newblock \bibinfo{title}{Classical time crystals}.
\newblock \emph{\bibinfo{journal}{Phys. Rev. Lett.}}
  \textbf{\bibinfo{volume}{109}}, \bibinfo{pages}{160402}
  (\bibinfo{year}{2012}).
  
\bibitem{isbn9780201657029}
\bibinfo{author}{Goldstein, H.}, \bibinfo{author}{Poole, C.~P.} \&
  \bibinfo{author}{Safko, J.~L.}
\newblock \emph{\bibinfo{title}{Classical mechanics}}
  (\bibinfo{publisher}{Addison-Wesley}, \bibinfo{year}{2002}).

  
\bibitem{luttinger1964}
\bibinfo{author}{Luttinger, J.~M.}
\newblock \bibinfo{title}{Theory of thermal transport coefficients}.
\newblock \emph{\bibinfo{journal}{Phys. Rev.}}
  \textbf{\bibinfo{volume}{135}}, \bibinfo{pages}{A1505}
  (\bibinfo{year}{1964}).

\bibitem{braginsky1977}
\bibinfo{author}{Braginsky, V.~B.}, \bibinfo{author}{Caves, C.~M.} \&
  \bibinfo{author}{Thorne, K.~S.}
\newblock \bibinfo{title}{Laboratory experiments to test relativistic gravity}.
\newblock \emph{\bibinfo{journal}{Phys. Rev. D}}
  \textbf{\bibinfo{volume}{15}}, \bibinfo{pages}{2047} (\bibinfo{year}{1977}).

\bibitem{ryu2012}
\bibinfo{author}{Ryu, S.}, \bibinfo{author}{Moore, J.~E.} \&
  \bibinfo{author}{Ludwig, A. W.~W.}
\newblock \bibinfo{title}{Electromagnetic and gravitational responses and
  anomalies in topological insulators and superconductors}.
\newblock \emph{\bibinfo{journal}{Phys. Rev. B}}
  \textbf{\bibinfo{volume}{85}}, \bibinfo{pages}{045104}
  (\bibinfo{year}{2012}).

\bibitem{qin2011}
\bibinfo{author}{Qin, T.}, \bibinfo{author}{Niu, Q.} \& \bibinfo{author}{Shi,
  J.}
\newblock \bibinfo{title}{Energy magnetization and the thermal hall effect}.
\newblock \emph{\bibinfo{journal}{Phys. Rev. Lett.}}
  \textbf{\bibinfo{volume}{107}}, \bibinfo{pages}{236601}
  (\bibinfo{year}{2011}) and supplemental text.

\bibitem{cooper1997}
\bibinfo{author}{Cooper, N.~R.}, \bibinfo{author}{Halperin, B.~I.} \&
  \bibinfo{author}{Ruzin, I.~M.}
\newblock \bibinfo{title}{Thermoelectric response of an interacting
  two-dimensional electron gas in a quantizing magnetic field}.
\newblock \emph{\bibinfo{journal}{Phys. Rev. B}}
  \textbf{\bibinfo{volume}{55}}, \bibinfo{pages}{2344} (\bibinfo{year}{1997}).
  
\bibitem{qin2012}
\bibinfo{author}{Qin, T.}, \bibinfo{author}{Zhou, J.} \& \bibinfo{author}{Shi,
  J.}
\newblock \bibinfo{title}{Berry curvature and the phonon hall effect}.
\newblock \emph{\bibinfo{journal}{Phys. Rev. B}}
  \textbf{\bibinfo{volume}{86}}, \bibinfo{pages}{104305}
(\bibinfo{year}{2012}).

\end{thebibliography}


\begin{addendum}
 \item We thank Jing-ning Zhang for pointing out to us an issue of original derivation.  This work is supported by 973 program of China (2009CB929101, 2012CB921304).
 \item[Competing Interests] The authors declare that they have no
competing financial interests.
 \item[Correspondence] Correspondence and and request for materials should be addressed to J.S. (email: junrenshi@pku.edu.cn).

\end{addendum}


\newpage

\end{document}